\documentclass[letterpaper]{article} 
\usepackage{aaai25}  
\nocopyright  
\usepackage{times}  
\usepackage{helvet}  
\usepackage{courier}  
\usepackage[hyphens]{url}  
\usepackage{graphicx} 
\usepackage{natbib}  
\usepackage{caption} 
\usepackage{amsmath}
\usepackage{booktabs}
\usepackage{multirow}
\usepackage{xcolor}
\usepackage{colortbl}
\usepackage{tikz}
\usetikzlibrary{arrows.meta}
\usepackage{enumitem}
\usepackage{makecell}
\usepackage{soul}

\usepackage{mdframed}
\usepackage{enumitem}

\newcommand{\answerYes}[1]{\textcolor{blue}{#1}}
\newcommand{\answerNo}[1]{\textcolor{teal}{#1}}
\newcommand{\answerNA}[1]{\textcolor{gray}{#1}}

\graphicspath{{figures/}}

\newcommand{\NumPosts}{7{,}945}
\newcommand{\NumComments}{288{,}277}
\newcommand{\NumHELPPosts}{5{,}526}
\newcommand{\NumGUESSPosts}{898}    
\newcommand{\NumGUESSPostsAI}{139}  
\newcommand{\NumGUESSPostsREAL}{190} 
\newcommand{\NumGUESSPostsGT}{329}  
\newcommand{\NumGUESSSentiment}{309} 
\newcommand{\NumHELPSentiment}{4{,}760}

\newcommand{\BotStartDate}{May 13, 2025}
\newcommand{\NumPostsAfterBot}{7{,}342}
\newcommand{\PostsPctAfterBot}{92.4\%}
\newcommand{\NumReasonComments}{11{,}774}
\newcommand{\NumAnnotated}{10{,}745}

\title{Humans Cannot Detect AI-Generated Media But Communities May --- For Now: Collaborative AI Detection in r/RealOrAI on Reddit}

\author{
    Tu\u{g}rulcan Elmas
}
\affiliations{
    The University of Edinburgh
}

\begin{document}

\maketitle

\begin{abstract}
We study human AI-detection behaviour at scale using a year of activity
from r/RealOrAI, a Reddit community where users collaboratively assess
whether visual media is real or AI-generated. The community is moderated by a bot that solicits verified labels from submitters of
self-challenging ``[GUESS]'' posts and publishes an aggregate
community prediction for each post, yielding naturalistic
ground truth at scale. Community detection accuracy reaches 72\% on [GUESS] posts with a systematic false-positive bias that intensifies over the year as the community's AI-suspicion grows.
Using a six-LLM ensemble validated against human-annotated
ground truth, we classify 10k reasoning-bearing comments along six cues covering perceptual features, context, consistency, AI knowledge, subject-matter expertise and provenance (tracing the media to its source).
Perceptual features (scene, visual artifacts, anatomy physics, lighting, behavior, text, audio) dominate reasoning (70\%) while
provenance verification is rarest (4\%) at the individual level but is amplified 4.3$\times$ in community
summaries, revealing aggregation as a reliability filter that selectively
surfaces diagnostic evidence. These findings reveal the limits of heuristic-based detection and show how online communities collectively navigate an increasingly contested information environment.



\end{abstract}

\section{Introduction}

AI-generated media now exhibit high perceptual fidelity, raising concerns about misinformation, trust erosion, and the reliability of human judgment in distinguishing authentic content from generated artifacts. While prior research has extensively studied human detection performance in controlled laboratory settings, considerably less is known about how people evaluate authenticity in the wild, particularly within large online communities where judgments are socially constructed and debated.

This paper examines collective AI-detection behavior in a naturalistic setting using data from the r/RealOrAI subreddit, a dedicated online community where users collaboratively assess whether visual media is real or AI-generated. Unlike experimental studies with fixed stimuli and explicit instructions, r/RealOrAI captures spontaneous judgments, uncertainty, disagreement, and justification through user comments and votes. Crucially, the community incorporates an automated moderation agent (RealOrAI-Bot) that provides verified ground-truth labels and aggregated community prediction (referred to as sentiment), enabling direct comparison between collective perception, expressed confidence, and actual authenticity.

Employing a longitudinal dataset spanning a year, we analyze the community responses through the strict binary format it enforces for all submissions: [HELP] posts, in which posters seek assistance in determining authenticity, and [GUESS] posts, in which posters already know if the media is real or AI and test the community's judgment. This enforced structure allows us to study both collective deliberation under genuine uncertainty and the community's detection performance when explicitly tested. By aligning individual and community responses with verified outcomes, we assess detection accuracy, reasoning cues, and how individual judgments aggregate into community consensus. Our research questions are as follows:

\begin{description}
  \item[RQ1:] To what extent can online communities accurately distinguish between real and AI-generated media in a naturalistic setting, and how does this detection performance change over time?
  \item[RQ2:] What reasoning cues do users rely on when judging the authenticity of media?
  \item[RQ3:] To what extent does aggregating individual judgments improve AI detection performance, and does community-level consensus emerge as a reliable signal where individual judgment fails?
\end{description}

Our results reframe AI detection from an individual perceptual challenge to a collective socio-technical process. We offer four primary contributions. First, to the best of our knowledge, we present the first large-scale empirical study of crowdsourced AI detection accuracy on Reddit, and the first to specifically examine the r/RealOrAI community in the wild with verified ground truth. Second, we show that aggregation of individually AI-biased judgments yields meaningful collective accuracy, 72.3\% on self-verified \texttt{[GUESS]} posts, though accuracy declines over the study period as the community's AI-suspicion bias grows. Third, we show that people over-rely on salient but undiscriminating perceptual heuristics while systematically underweighting highly diagnostic evidence. Finally, we characterise community aggregation as a reliability filter: diagnostic signals such as provenance are amplified in community summaries while ambiguous cues recede.


\section{Related Work}

The prevalence of AI-generated media worsens threats to information integrity. The rapid spread of realistic synthetic content on social platforms \cite{Yang2024FakeSocial,Ricker2024Twitter} not only facilitates misinformation \cite{Pote2024Coordinated,Gopalakrishnan2025Engagement} but also undermines baseline trust in authentic media, even when users correctly identify fabrications \cite{Vaccari2020Deepfakes}. Given the societal stakes of detection errors, finding reliable methods to verify synthetic content is an urgent priority. As we review below, neither automated systems nor isolated human judgment currently provide a reliable standalone defense against this threat, which necessitates our investigation into collective, community-level deliberation.

\subsection{Automated Detection of AI-Generated Media}

A substantial body of research focuses on automated detection of AI-generated media via supervised classifiers, statistical artifact analysis, and provenance techniques  (methods that trace a media item's origin or creation chain). While effective in controlled benchmarks, these methods struggle in real-world settings due to shifting generator distributions and evolving model artifacts. Previous work confirms a persistent generalisation problem: detectors degrade when generator distributions shift \cite{pei2024deepfake}, and performance degrades further when synthetic media are recompressed, resized, and reshared across social platforms \cite{Karageorgiou2024Evolution}.
Consequently, automated detection alone is insufficient for maintaining information integrity in open, user-driven online environments. 

Another effective automated approach is watermarking, verifiable signals embedded in the media at generation time. SynthID \cite{SynthID2024} and C2PA content credentials \cite{C2PA2024} shift detection from post-hoc inference to source attribution. In practice, these systems depend on widespread ecosystem adoption and remain vulnerable to metadata stripping during social media distribution, meaning many real-world items remain outside provenance rails \cite{croitoru2024deepfake}. Even when fingerprints are present, adversarial removal and forgery attacks succeed in over 80\% of white-box cases~\cite{Yao2026Smudged}.

These findings show that both algorithmic detection and provenance tracking remain fragile under real-world distribution shifts, highlighting the need to understand how human communities evaluate authenticity when automated systems fail and motivating our investigation of collective deliberation in r/RealOrAI.

\subsection{Human Ability to Identify AI-Generated Content}

Parallel to algorithmic approaches, a growing literature investigates human ability to distinguish between real and AI-generated media. Controlled experiments consistently show that human performance is limited, with accuracy often approaching chance as generative quality improves. Studies further indicate that individuals struggle to calibrate their confidence and often misinterpret visual cues \cite{Bray2023Testing,nightingale2022ai}.

Nightingale and Farid \citeyear{nightingale2022ai} show that AI-synthesised faces can be rated as both realistic and more trustworthy than real faces, challenging older ``uncanny valley'' assumptions. Groh et al.\ \citeyear{Groh2022Crowds} further show that humans and machine detectors make different errors and that machine-informed crowds can improve aggregate performance. Pratama et al.\ \citeyear{Pratama2025Perception} also argue that human perceptions of visual media differ systematically from those of machines and that human annotations are not yet replaceable by automated ones. Bray et al.\ \citeyear{Bray2023Testing} report limited deepfake-face detection accuracy and weak calibration between confidence and correctness. Chen et al.\ \citeyear{Chen2025TrainingFaces} compare three training strategies for detecting StyleGAN-synthesized faces, explicit instruction on visual artifacts, implicit exposure to the synthetic-face generation process, and a combination of both, and report that all three improve participants' detection accuracy and decision confidence relative to an untrained baseline. 


%

Crucially, while prior work shows that individuals struggle to reliably detect AI content in isolation, it typically relies on fixed stimuli and explicit evaluation tasks. This motivates our focus on an interactive social environment.

\subsection{Collective Deliberation and AI Detection}

Online communities provide naturalistic settings to study human reactions to controversies~\cite{Wang2026Grievance,bidewell2026gendered} including AI content~\cite{Yuce2026ChatGPTTeachers}. 
Collective deliberation about AI-generated content increasingly occurs in online communities, where authenticity judgments emerge through discussion rather than isolated evaluation. Lloyd et al.\ \citeyear{Lloyd2025Moderating} show that moderators often lack reliable automated tools and rely on community heuristics and labor-intensive review. This dynamic of crowdsourced assessment extends beyond Reddit: research on systems such as X's Community Notes demonstrates both the potential and limitations of collective deliberation in fact-checking and mitigating misleading content \cite{Sacha2022Crowdsourced,Chuai2024Community}. 

The closest empirical prior work examines collective reactions to suspected AI media in online communities. Matatov et al.\ \citeyear{Matatov2024ArtSubreddits} analyze 140 art-focused subreddits and report increasingly contested and often negative reactions around suspected AI use, including cases where suspicion exceeds available verification signals. Ha et al.\ \mbox{\citeyear{Ha2024Organic}} evaluate crowdworkers, professional artists, experts, and automated detectors on paired human-versus-AI art datasets and find distinct failure modes across groups. Roca et al.\ \citeyear{Roca2025HowGood} use a large-scale online ``Real or Not'' game dataset (approximately 287,000 independent image evaluations from over 12,500 participants), with aggregate accuracy only modestly above chance and substantial content-type heterogeneity. Their setting aggregates isolated judgements; r/RealOrAI instead captures \emph{collaborative deliberation}, in which users see and respond to one another's reasoning before community sentiment is summarised. Comparing these two modes and identifying when discursive aggregation outperforms independent aggregation is one of the questions we address.

Recent Reddit-focused work is especially relevant to our setting. Chauhan et al.\ \citeyear{Chauhan2025GenAIFakeNews} analyze repost cascades of misinformation and AI-generated images across five ideologically diverse subreddits and show broad cross-community propagation. Liu et al.\ \citeyear{Liu2026AILiteracy} analyze 122,000 Reddit conversations across 80 subreddits over three years and find that AI-literacy discourse is event-driven and release-sensitive. 


While prior studies analyze the volume and themes of community reactions to synthetic media, they leave a critical gap: evaluating the \textit{accuracy} of this collective detection against ground truth. As a community explicitly dedicated to the collective detection of AI-generated media, r/RealOrAI provides a unique environment to not only observe how crowds interactively build consensus, but to empirically quantify when and why the ``wisdom of crowds'' systematically fails to identify synthetic content in the wild.

\section{Data and Methods}
\label{sec:methods}

r/RealOrAI is a subreddit dedicated to the collaborative evaluation of visual media authenticity. Members submit images, videos, or audio clips and the community collectively decides whether each item is real or AI-generated. Community activity is mediated by the RealOrAI-Bot, an automated moderation agent that orchestrates ground-truth collection and aggregates community sentiment for each post. The community is governed by twelve explicit rules that shape submissions and community responses. Three affordances are particularly relevant to our study:

\noindent\textbf{Mandatory reasoning:}
Rule~3 requires that commenters who claim content is AI-generated provide an explanation for their judgment. As a result, the comment corpus contains large-scale, naturally occurring justifications. This enables systematic analysis of reasoning without experimental elicitation, as explanations are required by the platform.

\noindent\textbf{Built-in ground truth:} Rule~1 requires self-labeling posts into \texttt{[GUESS]} (submitter knows the answer and challenges the community to detect it) and \texttt{[HELP]} (submitter is genuinely uncertain and seeks the community's judgment). This was initially encoded as title prefixes (e.g., \texttt{[HELP] Is this AI?}). In March 2026, the community switched to Reddit \textit{flairs}, colored category labels displayed beneath a post's title. For \texttt{[GUESS]} posts, Rule~2 prohibits spoilers in the comment thread and routes the correct answer privately to the RealOrAI-Bot via DMs, which then publishes it after a delay. 
This delay is fixed at 12 hours, after which the bot reveals the ground-truth label and posts the aggregate sentiment score. As a result, all user responses are written before either signal is visible, preserving them as naturalistic detection judgments uncontaminated by ground-truth knowledge. 

Media type flairs (Photo, Video, Digital Art, Audio, Deepfake) were assigned by the submitter and enforced by moderators, but were discontinued in March 2026 and replaced with the \texttt{[GUESS]} \& \texttt{[HELP]} flairs described above.

\noindent\textbf{Content quality and evaluability:}
Rules~4 and~5 restrict submissions to actual media and exclude abstract art. Rule~7 further filters out memes and low-effort posts, requiring that every submission could genuinely be real or AI-generated. These constraints ensure the corpus consists of content where perceptual reasoning is both possible and non-trivial. The community also prohibits AI-generated text in comments and commercial data-collection activity, which helps reduce spam and low-effort participation. 

\begin{figure}[htbp]
\centering
\resizebox{\linewidth}{!}{%
\begin{tikzpicture}[
    x=1cm, y=1cm,
    every node/.style={font=\footnotesize},
    sbox/.style={
        rectangle, rounded corners=4pt,
        draw=gray!60, fill=gray!8,
        align=center, inner sep=6pt,
        text width=8.6cm, minimum height=1cm
    },
    gbox/.style={
        rectangle, rounded corners=4pt,
        draw=green!60!black, fill=green!8,
        align=center, inner sep=5pt,
        text width=3.9cm
    },
    hbox/.style={
        rectangle, rounded corners=4pt,
        draw=blue!60!black, fill=blue!8,
        align=center, inner sep=5pt,
        text width=3.9cm
    },
    botbox/.style={
        rectangle, rounded corners=4pt,
        draw=orange!70!black, fill=orange!8,
        align=center, inner sep=6pt,
        text width=8.6cm
    },
    arr/.style={-{Stealth[length=5pt,width=4pt]}, thick, gray!60},
    garr/.style={-{Stealth[length=5pt,width=4pt]}, thick, green!60!black},
    harr/.style={-{Stealth[length=5pt,width=4pt]}, thick, blue!60!black},
]

\node[sbox] (submit) at (0,0) {
    \textbf{Post Submitted to r/RealOrAI}\\[3pt]
    $N = \NumPostsAfterBot{}$ posts $\cdot$ \NumComments{} comments $\cdot$ May~2025--Apr~2026\\[2pt]
    \textit{Media type flair:}~Photo 39.2\% $\cdot$ Video 32.8\% $\cdot$ Digital Art 25.7\% $\cdot$ Audio 1.7\% $\cdot$ Deepfake 0.4\%
};

\node[gbox] (guess) at (-2.35,-2.2) {
    \textbf{[GUESS]}\\[2pt]
    $N=\NumGUESSPosts{}$ (10.2\%)\\
    Submitter knows the answer
};
\node[hbox] (help) at (2.35,-2.2) {
    \textbf{[HELP]}\\[2pt]
    $N=\NumHELPPosts{}$ (75.3\%)\\
    Submitter is uncertain
};

\node[sbox] (comments) at (0,-4.3) {
    \textbf{Community Commenting}~~(0--12\,h window)\\[2pt]
    Mandatory reasoning for AI claims $\cdot$ No spoilers $\cdot$ No AI-generated text\\[1pt]
    \NumReasonComments{} reasoning-bearing comments extracted from \NumComments{} total
};

\node[botbox] (bot) at (0,-6.55) {
    \textbf{RealOrAI-Bot pins comment at 12\,h}\\[3pt]
    Gemini~2.5~Flash reads up to 50 top comments\\[2pt]
    $\rightarrow$~\texttt{sentiment\_ai} $\in[0,1]$
    
    ~~(0 =  consensus real,~1 = consensus AI)\\
    $\rightarrow$~\texttt{brief\_reasoning}~~(natural-language summary)\\[1pt]
    Available for \NumGUESSSentiment{} [GUESS] and \NumHELPSentiment{} [HELP] posts;\\
    others lack enough comments within 12\,h to compute
};

\node[gbox] (gt_guess) at (-2.35,-9.1) {
    \textbf{[GUESS] Ground Truth}\\[2pt]
    Bot DMs OP at submission;\\
    \NumGUESSPostsGT{} of \NumGUESSPosts{} respond ($\approx$37\%)\\[2pt]
    285 w/ community sentiment
};
\node[hbox] (gt_help) at (2.35,-8.9) {
    \textbf{[HELP] Ground Truth}\\[2pt]
    \textit{No authoritative GT;}\\
    \textit{community sentiment only}
};

\draw[garr] (submit.south) to[out=250,in=90] (guess.north);
\draw[harr] (submit.south) to[out=290,in=90] (help.north);
\draw[garr] (guess.south)  to[out=270,in=165] (comments.west);
\draw[harr] (help.south)   to[out=270,in=15]  (comments.east);
\draw[arr]  (comments.south) -- (bot.north);
\draw[garr] (bot.south) to[out=250,in=90] (gt_guess.north);
\draw[harr] (bot.south) to[out=290,in=90] (gt_help.north);

\end{tikzpicture}}%
\caption{Lifecycle of a post on r/RealOrAI. \textcolor{green!60!black}{\textbf{Green}} = \texttt{[GUESS]} path (submitter knows the answer); \textcolor{blue!60!black}{\textbf{blue}} = \texttt{[HELP]} path (submitter is uncertain); grey = shared platform infrastructure. The RealOrAI-Bot pins a sentiment comment at 12\,h and, for \texttt{[GUESS]} posts, simultaneously DMs the submitter to request the verified label.}
\label{fig:platform_workflow}
\end{figure}
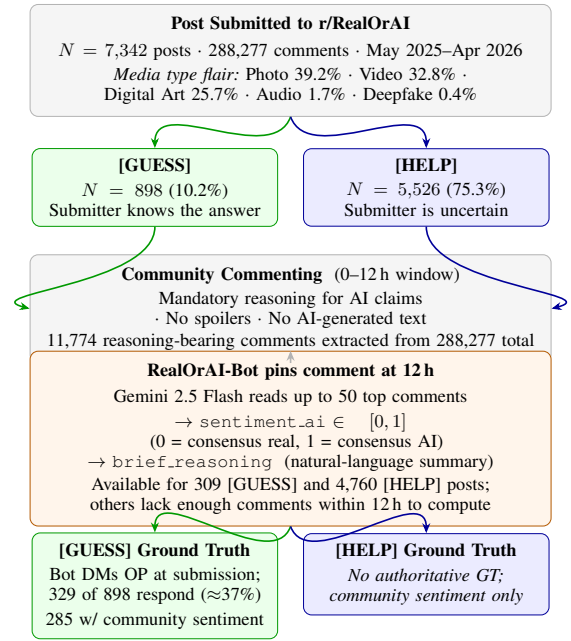

We collected \NumPosts{} posts and \NumComments{} associated comments spanning August 2022 to April 2026 using the Photon tool from~\cite{arcticshift2022}. The community grew slowly in its early years before undergoing rapid expansion: \PostsPctAfterBot{} of all posts (\NumPostsAfterBot{} of \NumPosts{}) were submitted after the RealOrAI-Bot was deployed on \BotStartDate{}, which also marks the onset of meaningful community activity. All quantitative analyses are restricted to this post-bot period. Figure~\ref{fig:platform_workflow} summarises the full post lifecycle, data flows, and ground-truth derivation. 

A notable mid-study change to the platform occurred in February 2026, when the bot was updated to append a \texttt{brief\_reasoning} field to its 12-hour comment, which is a short natural-language summary of the community's collective arguments for or against AI generation (available for 821 posts for both \texttt{[GUESS]} and \texttt{[HELP]}).  These 821 summaries constitute the \emph{brief-reasoning corpus} used in RQ3 to quantify individual and community differences in reasoning.

\subsection{LLM Classification Protocol}
\label{sec:llm-protocol}

Several tasks in this paper require classifying individual comments or posts into discrete categories: filtering meta comments, labelling comment verdicts, annotating reasoning cues, and assigning perceptual features. We apply a unified evaluation protocol to each task. Six large language models serve as independent classifiers: \textbf{Llama~3.3-70B-Instruct}, \textbf{Gemini~2.5~Flash}, \textbf{GPT-5.2}, \textbf{GPT-5-mini}, \textbf{Claude~Sonnet~4.6}, and \textbf{Claude~Haiku~4.5}. Each model receives a task-specific system prompt and produces structured output; full prompts are reproduced in Appendix~\ref{app:prompts}.

For each task, classifier selection is driven by comparison against a human-annotated ground truth. We evaluate each single model and all subsets of models under unweighted majority vote, exhaustively searched, and select the configuration with the highest F$_1$ score on the ground-truth set. Comments are batched 30 per request for all models except GPT-5-mini (10 per request, to fit its context window).


\section{RQ1: Community Detection Accuracy}
We compare community predictions against ground truth.

\noindent\textbf{Ground Truth:}
For \texttt{[GUESS]} posts, the RealOrAI-Bot publishes the verified label once the submitter responds to its DM (e.g., ``The answer is AI''). Of \NumGUESSPosts{} posts, \NumGUESSPostsGT{} include a label (\NumGUESSPostsAI{} AI-generated, \NumGUESSPostsREAL{} real), covering 37\% of posts.

\noindent\textbf{Community Prediction:}
The bot's 12-hour comment (Figure~\ref{fig:platform_workflow}) produces a continuous \textit{sentiment\_ai} score in $[0,1]$, where 0 indicates unanimous belief the content is real and 1 indicates unanimous belief it is AI-generated. We treat this as the community's predicted probability of AI generation and compare it to the ground truth. Sentiment scores are available for 309 of 898 \texttt{[GUESS]} posts (the rest lack enough comments within 12\,h); 285 of these also have a bot-verified label, yielding our evaluation subset (31.7\% of \texttt{[GUESS]} posts).

\noindent\textbf{Evaluation Metrics:}
We treat the bot-verified label as ground truth, binarise the bot's \textit{sentiment\_ai} score at 0.5 and evaluate it using accuracy, precision, recall, and F1 score.

To capture the degree of misalignment between collective belief and verified authenticity, we define the \emph{community prediction error} as $E = | \textit{sentiment\_ai} - y |$, where $y \in \{0,1\}$ is the ground-truth label (0 for real, 1 for AI). $E$ ranges from 0 (community perfectly calibrated) to 1 (maximum miscalibration), and its mean over any sample equals the mean absolute error (MAE) of the probabilistic prediction against the binary outcome. 
Temporal trends are visualised using rolling averages with linear trend overlays. 

\subsection{Detection Performance}
\label{sec:rq1-help-eval}

\noindent\textbf{[GUESS] posts:} We evaluate community detection performance on the \texttt{[GUESS]} posts for which both a bot-verified label and a community sentiment score are available ($N = 285$ posts; 119 AI-labelled, 166 REAL-labelled). Community sentiment is binarised at 0.5. Overall accuracy is \textbf{72.3\%} (Table~\ref{tab:flair_stats}). The community exhibits a systematic false-positive bias: 55 real posts are flagged AI by the community versus only 24 AI posts missed  (a 2.3:1 ratio), reflected in AI recall (0.798) substantially exceeding AI precision (0.633). The community shows a prior bias toward attributing posts as AI. Average prediction error is high: the mean community prediction error is $\bar{E} = 0.327$, i.e., the community's mean probabilistic estimate sits $\approx$33 percentage points away from the binary ground truth, and 25.3\% of posts have $E > 0.5$, meaning the community's collective sentiment strongly favours the wrong answer.

\noindent\textbf{Media Type:} Detection performance varies substantially by media type (Table~\ref{tab:flair_stats}). \textbf{Video has the highest F\textsubscript{1}} (76.6\%), with a clear signal separation between bot-verified AI videos (mean community AI-sentiment 78.8\%) and bot-verified REAL videos (33.5\%), and only 3 false negatives versus 8 false positives. \textbf{Digital Art follows} (F\textsubscript{1} = 71.0\%), which may be because AI-generated art exhibits the most salient stylistic artefacts and users bring stronger priors about what AI art looks like; with $N=38$ this should be treated as suggestive. \textbf{Photo is hardest} (F\textsubscript{1} = 69.9\%), contributing the most false positives (34 false positives vs.\ 16 false negatives).

\begin{table}[htbp]
\centering\small
\begin{tabular}{@{}lrrrrrrr@{}}
\toprule
\textbf{Flair} & \textbf{AI\%} & \textbf{Total} & \textbf{N} & \textbf{$n_{\text{AI}}$} & \textbf{Prec} & \textbf{Rec} & \textbf{F\textsubscript{1}} \\
\midrule
Photo      & 60.1 & 474 & 179 &  74 & 63.0 & 78.4 & 69.9 \\
Video      & 43.4 & 201 &  44 &  21 & 69.2 & 85.7 & 76.6 \\
Dig.\ Art  & 61.5 & 116 &  38 &  15 & 68.8 & 73.3 & 71.0 \\
No Flair   & 56.1 & 104 &  24 &   9 & 50.0 & 88.9 & 64.0 \\
\midrule
Total      & 54.8 & 898 & 285 & 119 & 63.3 & 79.8 & 70.6 \\
\bottomrule
\end{tabular}
\caption{AI-class metrics by flair on \texttt{[GUESS]} posts. AI\% = mean community sentiment across all posts of that flair with a sentiment score (\texttt{[GUESS]}+\texttt{[HELP]} combined). Total = all \texttt{[GUESS]} posts of that flair. N = evaluation subset (posts with both bot-verified label and sentiment); $n_{\text{AI}}$ = AI-labeled posts within N. Audio (123 posts; 2 \texttt{[GUESS]}) and Deepfake (31; 1 \texttt{[GUESS]}) are omitted: too few verified cases.}
\label{tab:flair_stats}
\label{tab:guess_vs_help}
\end{table}

\subsection{Temporal Trends in Detection Performance}
\label{sec:rq1-temporal}

\begin{figure*}[!ht]
\centering
\includegraphics[width=\linewidth]{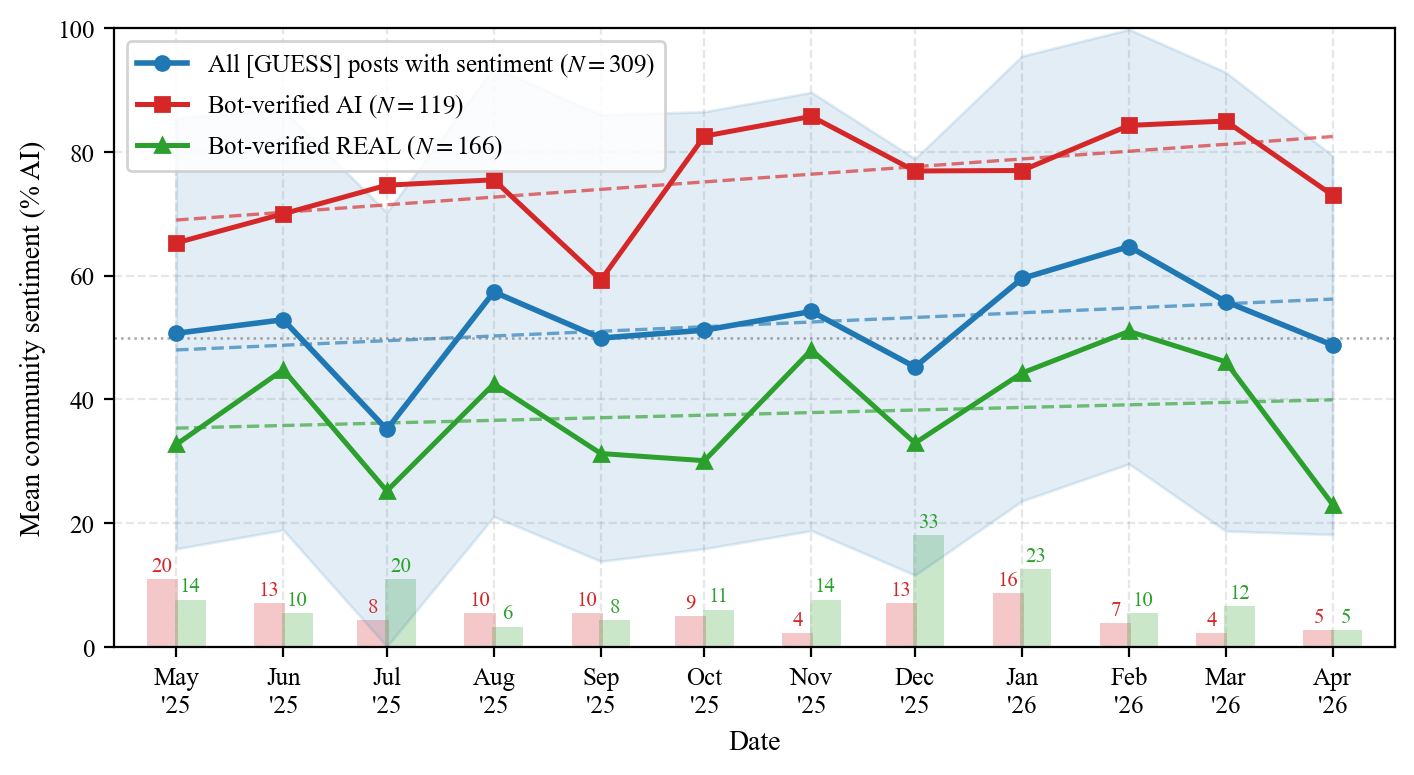}
\caption{{Monthly mean community sentiment on bot-verified \texttt{[GUESS]} posts, decomposed by ground-truth label: all \texttt{[GUESS]} posts with sentiment (blue), bot-verified AI (red), bot-verified REAL (green). Bars show monthly AI/REAL post counts; dashed lines show per-series linear trends.}}
\label{fig:sentiment_time}
\end{figure*}

Figure~\ref{fig:sentiment_time} tracks monthly mean community sentiment on bot-verified \texttt{[GUESS]} posts over the full study window. AI posts average 73.7\% mean sentiment and REAL posts 37.4\%, consistent with the aggregate metrics in Table~\ref{tab:flair_stats}. Visual inspection suggests a positive linear trend in mean sentiment for both AI and REAL posts. Monthly REAL sentiment reaches 46--51\% in November~2025 and February--March~2026, approaching the $0.5$ binarisation threshold.

To estimate sentiment trends, we employ post-level OLS regressions on days-since-collection-start, separately by ground-truth label and pooled with an interaction term. Accuracy and false-positive-rate trends are estimated via post-level logistic regressions. Sentiment on AI posts rises significantly (slope $+1.58$\,pp/month, $p=0.042$), while REAL sentiment shows a positive but non-significant trend ($+1.28$\,pp/month, $p=0.085$), with no detectable difference in slopes across classes (interaction $p=0.78$). The community's growing AI-suspicion thus appears \emph{uniform} across both classes rather than \emph{selective} to AI content. The behavioural consequence is asymmetric across classes: on REAL posts the odds of misclassification as AI grow significantly over the study window (log-odds $+0.15$/month, $p=0.007$), while on AI posts accuracy improves at a comparable rate (log-odds $+0.13$/month, $p=0.056$). These two effects approximately cancel at the aggregate, so overall accuracy does not significantly decline ($p=0.22$). These results suggest that the community is becoming increasingly inclined to misclassify authentic content as AI-generated. This is with a population-level shift in priors as generative-AI content becomes more prevalent.


%

\section{RQ2: Reasoning Strategies \& Cues}
\label{sec:rq2}

We use LLM-based classification to identify reasoning cues of users arguing why a particular media is real or AI.


\noindent\textbf{Pipeline:} The classification pipeline has four layers: (1) a \textsc{Meta} filter removing verdict-free comments; (2) a five-class \textit{verdict} label (\textsc{AI}, \textsc{Real}, \textsc{Other}, \textsc{Both}, \textsc{Unknown}); (3) six binary \textit{reasoning cues} (\textit{perceptual feature}, \textit{consistency}, \textit{provenance}, \textit{context reasoning}, \textit{AI knowledge}, \textit{subject knowledge}); and (4) \textit{perceptual subcategories} applied only to comments that fired the perceptual feature cue, with all others receiving an empty set. Each layer is defined in the subsections below. Each layer uses the LLM classification protocol (Section~\ref{sec:llm-protocol}); classifier selection for each layer is driven by comparison against a human-annotated ground truth. (See Appendix~\ref{app:prompt-meta} for the system prompts)

The schema was developed using a combination of (1) manual reading of a sample of 100 posts; (2) zero-shot prompting of multiple LLMs (Gemini 3, Claude Opus, GPT 5.3), where each model was asked on the fly to explain comment-level judgments (e.g., “Why does this user think the content is AI or real?”) without additional guidance; and (3) using Google NotebookLM’s Mind Map method to extract themes from reasoning-focused comments. Infrequent cues (e.g., ``artist workflow") are removed or merged into broader ones and near-co-occurring semantically similar pairs (e.g., ``Lighting" and ``Geometry") are merged.

\noindent\textbf{Data Sampling:} As our objective is to obtain reasoning-bearing comments, we adopt a strategic data sampling approach to maximize the rate of such comments and avoid noise. First, we only keep top-level comments (60\%) as the replies to other comments are predominantly social exchanges (e.g., ``agreed'') or continuations of an earlier argument. Of top-level comments, 67\% contain an explicit verdict marker (e.g.\ \textit{AI}, \textit{real}, \textit{fake}). From this pool, we collect comments that contain an explicit causal connective after a verdict (\textit{because}, \textit{since}, \textit{due to}, \textit{as}), capturing sentences of the form ``\textit{[verdict] [connective] [reason]}''. Patterns case-insensitively match anywhere in the comment text rather than being anchored to sentence starts, and near-duplicate comments are removed after extraction. We identify \NumReasonComments{} reasoning-bearing comments (15\% of the top comments with verdicts). This method trades recall for precision and substantially reduces the cost of independently classifying each comment with six LLMs.

%

\noindent\textbf{Ground truth:} For LLM classifier selection, we constructed a 150-comment human-annotated ground-truth sample. Comments were sampled stratified by model disagreement to prioritise difficult and ambiguous cases: for each of six dimensions in the initial classification schema (verdict, perceptual feature, consistency, provenance, subject knowledge, and AI knowledge) 25 rows were selected from the pool in which three initial LLM classifiers (Llama~3.3-70B, Gemini~2.5~Flash, and Claude~Sonnet~4.6) did not unanimously agree, yielding 150 non-overlapping comments ($6 \times 25$). For the perceptual feature dimension, when multiple comments had identical three-way disagreement scores, ties were broken by selecting comments from whichever subcategory had the fewest representatives already in the sample; the resulting GT covers all nine perceptual subcategories (per-subcategory $n$ in Table~\ref{tab:reasoning_and_subcats}). An expert annotator labelled the sample against this initial schema and a student annotator then independently reannotated all 150 comments against the updated schema, providing ground truth for all dimensions including the new \textit{context reasoning} and \textsc{Meta filter}. Final labels were established through crosschecking and discussion of disagreements. Per-flag agreement statistics are reported in the respective subsections below. 

\noindent\textbf{Meta comment filtering:}
A subset of reasoning-bearing comments are \textsc{Meta}: they address the subreddit, other users, or general AI discourse in the abstract, and contribute no verdict signal about the specific media. Assigning verdict or perceptual-cue labels to such comments introduces spurious annotations (e.g.\ a comment generically mentioning ``hands and eyes'' as AI cues incorrectly fires the perceptual cue flag without referring to any specific content). We therefore apply a binary \textsc{Meta} filter before all downstream annotation. Both annotators agreed on 147 of 150 \textsc{Meta} labels ($\kappa = 0.86$) before reconciliation. GPT-5.2 achieves the best classification performance against the human ground truth (F1~$= 0.857$, $\kappa = 0.846$) and is used for full-corpus annotation of \NumReasonComments{} comments. Approximately 8.7\% of comments are flagged as \textsc{Meta} and excluded from all subsequent layers, yielding \NumAnnotated{} content-bearing comments. (See Appendix~\ref{app:prompt-meta} for the LLM prompt)

\label{sec:verdict-methods}
\noindent\textbf{Verdict:}
Each comment receives one of five \textit{verdict} labels: \textsc{AI}, \textsc{Real}, \textsc{Other} (human-made but non-authentic: CGI, Photoshop, named manipulation), \textsc{Both} (argues both directions), or \textsc{Unknown} (no directional claim). Inter-annotator agreement on the ground-truth sample is $\kappa = 0.60$ (5-class) / $\kappa = 0.79$ (binary; Table~\ref{tab:reasoning_and_subcats}). Classifier selection follows the protocol in Section~\ref{sec:llm-protocol}; a \textbf{4-model majority vote} (GPT-5.2, GPT-5-mini, Haiku~4.5, Gemini~2.5~Flash) achieves the best performance ($\kappa = 0.675$, 5-class), outperforming the best single model (GPT-5.2, $\kappa = 0.641$), and is used for full-corpus annotation. 2-way ties, occurring in 4.5\% of full-corpus rows, are broken by priority order that follow corpus label frequency: AI\,$\succ$\,REAL\,$\succ$\,BOTH\,$\succ$\,OTHER\,$\succ$\,UNKNOWN

\noindent\textbf{Reasoning cue:}
Beyond the verdict, each comment is annotated with six binary \emph{reasoning cues} characterising \emph{how} the commenter constructs their argument. Cues are independent and may co-occur in the same comment. The six cues are:

\begin{itemize}[noitemsep]
  \item \textit{Perceptual feature:} names a concrete perceptual detail: a body part, texture, lighting condition, shadow, motion characteristic, or rendering artefact.
  \item \textit{Consistency:} checks whether some element remains stable across frames (for video) or across images that should be mutually consistent (e.g., ``... the guy on the background appears and disappears between frames'').
  \item \textit{Provenance:} actively verifies origin via an external source, such as a reverse image search, a social-media lookup, or an archive check
  \item \textit{Context reasoning:} reasons about whether the scenario depicted is plausible in reality, or about who would produce this content and why (e.g., ``... this type of event would be major news, why is there no other footage?'').
  \item \textit{AI knowledge:} invokes specific knowledge \emph{about AI}: a named model or tool (e.g., Midjourney, Sora), an AI capability timeline argument, or a failure mode (e.g., ``AI always messes up hands'').
  \item \textit{Subject knowledge:} invokes first-person expertise in the \emph{subject matter} depicted (e.g., ``as a nurse, those vitals are wrong'', ``I own horses - they don't move like that''). Unlike the other cues, this captures the \emph{source of authority} behind the claim rather than the form of the reasoning.
\end{itemize}

Classifier selection follows the LLM classification protocol from Section~\ref{sec:llm-protocol}; Table~\ref{tab:reasoning_and_subcats} reports the best classifier and performance against the human ground truth for each dimension, alongside empirical prevalence across the corpus.

\noindent\textbf{Perceptual Feature classification:}
A fourth classification layer labels \emph{which perceptual domain} the reasoning concerns, applied only to comments where the \textit{perceptual feature} flag is set. The nine subcategories and their descriptions are listed in Table~\ref{tab:reasoning_and_subcats}. Comments may be annotated with multiple perceptual features. Comments where the reasoning is purely strategic (e.g., a provenance check with no perceptual content) receive an empty set. Global absence-of-wrongness statements (``nothing looks off'', ``nothing distorted'', ``the focus looks fine'') leave \textit{perceptual\_feature}~=~false and receive no perceptual cue subcategory: the named feature must be present in the comment; its absence cannot substitute. 

After the initial LLM classification test, we observed low Anatomy classification performance despite the category being relatively easy to annotate. Inspection showed that LLMs consistently labelled Anatomy-related cues only as Visual Artifacts (e.g., rendering errors with fingers), likely because anatomy errors are a subset of visual artifacts. We therefore performed an additional classification step to separate Anatomy-related cues from other Visual Artifacts, improving the Anatomy F1 score from 0.50 to 0.58. (See Appendix~\ref{app:prompt-v12} for the prompt.)


\subsection{Results}

\noindent\textbf{Verdict distribution:} Across \NumAnnotated{} classified comments, \textsc{AI} is the plurality verdict (5{,}702; 53.1\%), with \textsc{Real} (3{,}847; 35.8\%) the second most common. Approximately 4.2\% of comments (456) are classified as \textsc{Other}: arguments about human-made manipulation that binary schemas would misattribute, potentially inflating apparent AI detection rates. \textsc{Unknown} (471; 4.4\%) and \textsc{Both} (269; 2.5\%) reflect genuine uncertainty or explicitly hedged reasoning.


\noindent\textbf{Reasoning cues and perceptual subcategories:} Table~\ref{tab:reasoning_and_subcats} reports cue prevalence and perceptual subcategory rates with two complementary metrics: \textbf{P(AI\textbar{}cue)} is the AI-verdict rate among comments using that cue (how strongly invoking the cue is associated with an AI verdict); \textbf{P(cue\textbar{}AI)} is the rate of the cue among AI-verdict comments (selectivity: how often AI-verdict commenters rely on this cue). The overall AI base rate is 53.1\%; cues with P(AI\textbar{}cue) above this are AI-associated, below it are REAL-associated.

\begin{table*}[!ht]
\centering\footnotesize
\resizebox{\linewidth}{!}{%
\begin{tabular}{@{}lp{4.7cm}clr@{\,\,}r@{\,\,}r@{\,\,}r@{\hspace{6pt}}r@{\,\,}r@{\,\,}r@{}}
\toprule
 & & & & \multicolumn{4}{c}{\textbf{Classification}} & \multicolumn{3}{c}{\textbf{Prevalence}} \\
\cmidrule(lr){5-8} \cmidrule(l){9-11}
\textbf{Cue} & \textbf{Description} & \textbf{$\kappa$} & \textbf{Best Classifier} & \textbf{$n$} & \textbf{P} & \textbf{R} & \textbf{F\textsubscript{1}} & \textbf{\%} & \textbf{P(AI\textbar{}c)} & \textbf{P(c\textbar{}AI)} \\
\midrule
Meta filter & Non-content-directed comments & 0.86 & G5$^\ast$ & 10 & 0.82 & 0.90 & 0.86 & -- & -- & -- \\
Verdict & AI, REAL, Other, Both, Unknown & 0.60 & G5\,+\,G5m\,+\,H\,+\,Gem$^\dagger$ & 140$^\ddagger$ & 0.59 & 0.67 & 0.60 & -- & -- & -- \\
\hspace{1em}(binary) & Binary AI vs.\ non-AI & 0.79 & G5\,+\,G5m\,+\,H\,+\,Gem$^\dagger$ & 140$^\ddagger$ & 0.92 & 0.92 & 0.92 & -- & -- & -- \\
\midrule
Perceptual Feature & Names a concrete perceptual detail & 0.60 & G5m $+$ So & 87 & 0.82 & 0.86 & 0.84 & 70.5 & 58.5 & 77.8 \\
Context Reasoning & Argues from content's plausibility & 0.31 & Gem $+$ G5m $+$ So $+$ H & 27 & 0.61 & 0.52 & 0.56 & 20.7 & 57.2 & 22.3 \\
Consistency & Checks cross-frame stability & 0.67 & Gem $+$ G5 & 12 & 0.92 & 0.92 & 0.92 & 10.4 & 48.3 & 9.5 \\
AI Knowledge & Invokes AI failure modes or tools & 0.47 & Ll $+$ Gem $+$ G5 $+$ So & 22 & 0.94 & 0.73 & 0.82 & 8.1 & 37.8 & 5.8 \\
Subject Knowledge & Claims subject-matter expertise & 0.44 & G5m & 8 & 1.00 & 0.75 & 0.86 & 4.4 & 34.9 & 2.9 \\
Provenance & Verifies origin via search/archive & 0.76 & Ll $+$ Gem $+$ H & 20 & 0.93 & 0.70 & 0.80 & 3.9 & 31.7 & 2.4 \\
\midrule
Visual Artifacts & Blurring, warping, rendering failures & 0.32 & H $+$ Ll & 13 & 0.53 & 0.69 & 0.60 & 25.6 & 65.9 & 22.4 \\
Scene \& Env. & Background objects, impossible scenes, spatial coherence & 0.54 & G5m $+$ Gem $+$ Ll & 19 & 0.74 & 0.74 & 0.74 & 22.0 & 64.3 & 18.8 \\
Physics \& Motion & Unrealistic movement, gravity & 0.55 & Gem $+$ H & 13 & 0.90 & 0.69 & 0.78 & 19.1 & 62.2 & 15.8 \\
Behavior \& Psych. & Facial expressions, gaze, body language, emotional plausibility & 0.61 & So $+$ Ll & 10 & 0.77 & 1.00 & 0.87 & 13.3 & 59.0 & 10.4 \\
Text \& Details & Readable text, signs, fine detail & 0.39 & Gem & 14 & 0.75 & 0.64 & 0.69 & 13.2 & 53.1 & 9.3 \\
Anatomy & Faces, hands, fingers, limbs & 0.48 & Gem $+$ G5m $+$ H & 16 & 0.50 & 0.69 & 0.58 & 12.8 & 66.3 & 15.9 \\
Lighting \& Geom. & Shadows, reflections, perspective & 0.52 & G5m & 19 & 0.86 & 0.63 & 0.73 & 9.0 & 53.4 & 6.4 \\
Audio & Voice quality, AV sync issues & 0.79 & G5 & 8 & 0.89 & 1.00 & 0.94 & 4.2 & 74.3 & 4.2 \\
Imperfections$^\S$
& Film grain, noise, natural wear & 0.44 & Gem & 4 & 1.00 & 0.75 & 0.86 & 1.4 & 11.2 & 0.2 \\
\bottomrule
\end{tabular}}
\caption{LLM classification performance and empirical prevalence across \NumAnnotated{} comments (after meta filtering).\\
\textbf{$\kappa$}: Cohen's $\kappa$ between two human annotators on the 150-row ground-truth sample (binary for cues/subcats; 5-class and binary for verdict). $n$ = GT positive count (binary tasks) or total evaluation rows (verdict$^\ddagger$). Verdict P/R/F$_1$ are macro-averaged across 5 classes; binary verdict P/R/F$_1$ are macro-averaged across 2 classes (AI\,/\,non-AI); \\
\textbf{\%}: prevalence among all comments (cues) or among perceptual-feature comments (subcats).
\textbf{P(AI\textbar{}c)}: fraction of cue users concluding AI (vs 53.1\% base rate). \textbf{P(c\textbar{}AI)}: fraction of AI-verdict comments using this cue.\\
\textbf{Gem}=Gemini~2.5~Flash, \textbf{G5}=GPT-5.2, \textbf{G5m}=GPT-5-mini, \textbf{H}=Haiku~4.5, \textbf{Ll}=Llama~3.3-70B, \textbf{So}=Sonnet~4.6.\\
$^\dagger$~Majority vote; ties broken by AI$\succ$REAL$\succ$BOTH$\succ$OTHER$\succ$UNKNOWN. $^\ddagger$~Verdict $n$ = total non-meta rows; no single positive class. $^\S$~\textit{Imperfections} is asymmetric by design: reserved for arguments supporting authenticity; the inverse ``too smooth'' claim is coded as Visual Artifacts.}

\label{tab:reasoning_and_subcats}
\end{table*}


Cue prevalence is heavily skewed: \textbf{perceptual feature citation} dominates at 70.5\%  (77.8\% of all AI-verdict comments invoke at least one concrete perceptual detail), while the remaining cues each appear in fewer than 21\% of comments. \textbf{Context reasoning} (20.7\%) is the only other commonly used cue; \textbf{consistency checking} (10.4\%), \textbf{AI knowledge} (8.1\%), \textbf{subject knowledge} (4.4\%), \textbf{provenance} (3.9\%) form a sparse long tail, despite their reliability (See RQ3). The latter three are all substantially more common in REAL-verdict than AI-verdict comments (provenance: 5.8\% vs 2.4\%; subject knowledge: 7.0\% vs 2.9\%; AI knowledge: 12.1\% vs 5.8\%). The AI knowledge finding is counterintuitive: commenters who invoke AI-specific failure mode knowledge are in practice \emph{more} likely to conclude REAL, suggesting that knowledgeable users may be more likely to identify false positives. 


The perceptual subcategories most strongly associated with AI verdicts are \textbf{Audio} (74.3\% of audio-citing comments conclude AI), \textbf{Anatomy} (66.3\%), and \textbf{Visual Artifacts} (65.9\%): when commenters invoke voice quality, anatomical distortions, or rendering failures, they are substantially more likely than the 53.1\% base rate to call the content AI-generated. The \textbf{Audio} cue exhibits a striking asymmetry: despite being one of the least prevalent subcategories (4.2\%, naturally bounded by the fraction of video/audio posts in the corpus), it strongly drives AI conclusions. Because current generative models struggle with audio-video synchronization and realistic voice prosody, commenters heavily rely on these specific temporal failures to argue for artificiality, rarely citing audio to prove authenticity. Similarly, \textbf{Anatomy} is predominantly invoked as an AI accusation (66.3\% vs.\ the 53.1\% base rate): because humans possess hyper-tuned perceptual sensitivity to uncanny-valley effects and generative models historically struggle with structural limb coherence (e.g., finger counts), anatomical citations more often signal artificiality than authenticity.

\textbf{Visual Artifacts} is the most common perceptual subcategory (25.6\%) and strongly associated with AI verdicts, making it the dominant mode of AI reasoning overall. In contrast, \textbf{Text \& Details} (53.1\%) and \textbf{Lighting \& Geometry} (53.4\%) are near the base rate: verdict-neutral cues used in both directions in roughly equal numbers (e.g., ``the text is garbled, so it's AI'' vs.\ ``the background text is perfectly legible, so it must be real''). \textbf{Consistency} is similarly bidirectional: at a 48.3\% AI-verdict rate, commenters invoke it equally to argue that shifting backgrounds betray AI generation and that stable, long-take physics prove authenticity. Finally, \textbf{Imperfections} has the lowest AI-verdict association (11.2\%) by design: the subcategory is reserved for arguments \emph{supporting} authenticity (e.g., film grain, compression noise, natural wear), while the inverse ``too smooth'' claim is coded as Visual Artifacts. These features reveal a ``fatal flaw'' vs.\ ``general vibe'' spectrum: Audio and Anatomy are rare but represent hard violations that leave little room for doubt; Visual Artifacts and Scene \& Environment are cited far more often (25.6\% and 22.0\%) but are diffuse, since authentic media can appear strangely composed or heavily filtered. No single feature dominates: even Visual Artifacts appears in only 22.4\% of AI-verdict comments, followed by Scene \& Environment (18.8\%) and Physics \& Motion (15.8\%), reflecting the diversity of generative failure modes across media types.

\section{RQ3: Individual vs Community}
\label{sec:rq3}
We analyse detection performance at three levels of aggregation (individual reasoning comments, aggregated reasoning comments, and full community sentiment from RQ1). We also examine how reasoning cues are amplified or suppressed as judgements move from individual comments to community-level summaries.

\noindent\textbf{Setup:} We identify 685 reasoning comments on 205 bot-verified \texttt{[GUESS]} posts (75 AI-labelled, 130 REAL-labelled). Each comment's ensemble verdict (Section~\ref{sec:verdict-methods}) provides an \emph{individual human prediction}. We retain only AI and REAL ensemble verdicts, discarding BOTH, OTHER, and UNKNOWN. To assess aggregated reasoning, we aggregate comments to the same post by majority vote: the post is predicted AI iff a strict majority of its active verdicts is AI. We also compute community prediction provided by the RealOrAI-Bot for the same post pool and all posts (same as RQ1).
Table~\ref{tab:individual_vs_community} reports results for the matched set, posts with both active reasoning verdicts and bot sentiment: $N=157$ and all \texttt{[GUESS]} posts with community sentiment ($N=285$).

\begin{table}[htbp]
\centering\footnotesize
\setlength{\tabcolsep}{3pt}
\renewcommand{\arraystretch}{1.0}
\begin{tabular}{@{}lrrrrrr@{}}
\toprule
\textbf{Method} & \textbf{N} & \textbf{Acc} & \textbf{P} & \textbf{R} & \textbf{F$_1$} & \textbf{AUC} \\
\midrule
\multicolumn{7}{l}{\textit{\textsc{[guess]}: bot-verified (75 AI, 130 REAL)}} \\[1pt]
\quad Individual reasoning  & 594 & 0.66 & 0.62 & 0.78 & 0.69 & -- \\
\quad Aggregated reasoning  & 157 & 0.64 & 0.54 & 0.73 & 0.62 & 0.69 \\
\quad Community (matched)   & 157 & 0.72 & 0.61 & 0.79 & 0.69 & 0.81 \\
\quad Community (all)       & 285 & 0.72 & 0.63 & 0.80 & 0.71 & 0.80 \\
\midrule
\multicolumn{7}{l}{\textit{By reasoning-comment count per post}} \\[1pt]
\quad 1 comment         &  92 & 0.65 & 0.46 & 0.62 & 0.53 & 0.64 \\
\quad 2 comments$^*$    &  26 & 0.65 & 0.63 & 0.77 & 0.69 & 0.65 \\
\quad 3 comments        &  20 & 0.45 & 0.31 & 0.67 & 0.42 & 0.71 \\
\quad $\geq$4 comments  &  27 & 0.85 & 0.81 & 0.93 & 0.87 & 0.92 \\
\bottomrule
\end{tabular}
\caption{AI-detection performance at multiple levels of aggregation on \textsc{[Guess]} posts. Positive class\,=\,AI. Comment-count rows pool the 165 posts with $\geq$1 active reasoning verdict (8 more than the 157-post matched set, which additionally requires bot sentiment). $^*$Excludes the 19 of 45 posts where the two active verdicts tie; tie-breaking toward REAL yields 0.67, toward AI 0.51.}
\label{tab:individual_vs_community}
\end{table}

\noindent\textbf{Individual reasoning comments are noisy and systematically AI-biased:} 594 of 685 reasoning comments (86.7\%) produce an active verdict with individual accuracy of 0.66 against bot-verified labels which is below the 0.69 always-REAL baseline given the 130/75 (REAL/AI) composition of the GT-matched set. Commenters' systematic AI-detection bias generates more false positives than missed positives at the individual level.

\noindent\textbf{Aggregation improves accuracy only with enough explicit reasoners:} Pooling all active reasoning verdicts within each post yields Acc = 0.64, lower than individual comment accuracy. However this is a structural artefact: as Table~\ref{tab:individual_vs_community} shows, the accuracy is low for posts with two and three reasoning comments. Two-comment posts introduce a structural tie problem: 19 of 45 posts produce an exact 50/50 split. Ties are uninformative by construction and excluded; the table reports accuracy only for unanimous 2-comment posts ($N=26$ \texttt{[GUESS]}), which score 0.654. Three-comment posts achieve only 0.450 accuracy ($N=20$). The AI-detection bias of individual commenters easily achieves a 2-1 majority on REAL posts: 11 of 20 posts are misclassified, and 9 of those are REAL posts called AI. Meanwhile, $\geq$4 comments accuracy is high (0.85), showing that aggregating sufficient number of explicit reasoners improves accuracy (2 ties are discarded).

\noindent\textbf{Full community sentiment is the strongest predictor:} On the matched \texttt{[GUESS]} set ($N=157$), bot sentiment reaches Acc = 0.72, F$_1$ = 0.69, AUC = 0.81; on all 285 \texttt{[GUESS]} posts with sentiment, performance is essentially the same (Acc = 0.72, F$_1$ = 0.71, AUC = 0.80), indicating that restricting to posts with reasoning comments does not bias the community-sentiment metrics.

\subsection{Community-level Cue Amplification} To characterise the community level reasoning extensively, we employ the 821 brief-reasoning summaries provided by the bot. The summaries cover 774 \texttt{[HELP]} posts and 47 \texttt{[GUESS]} posts. Thus, we do not evaluate their detection performance due to the low number of \texttt{[GUESS]} posts and report results by comparing community sentiment to individual sentiment. We use the same best-classifer ensembles from RQ2 (Table~\ref{tab:reasoning_and_subcats}) to classify the reasoning flag and perceptual cues mentioned in the brief reasonings. Table~\ref{tab:brief_reasoning_amplification} reports how brief-reasoning cue rates compare to the individual-comment rates from RQ2. 

Two divergences stand out at the flag level. First, \textbf{provenance} is massively amplified at the community level: whereas only 3.9\% of individual reasoning comments cite an external source, 20.6\% of community summaries include provenance evidence. This explains why aggregating explicit reasoning comments alone yields only 0.643 accuracy on \texttt{[GUESS]}: reasoning comments have low provenance density, but the full thread often contains one definitive source comment that drives the collective verdict. When that source is present, it dominates the community summary even though it appeared in only a fraction of individual comments.  The low individual rate likely reflects \texttt{[GUESS]}'s platform design: Rule~2 prohibits posting answers in the comment thread, which would typically include source URLs that reveal the verdict, plausibly suppressing direct provenance citations. This formal rule is consistent with a broader game-frame norm in which looking up the source may be treated as outside the spirit of the challenge.

Second, \textbf{context reasoning} undergoes a striking directional reversal. Among individual comments, context reasoning is weakly associated with AI verdict (P(AI$|$cue) = 57.2\%). Among community summaries, the same flag is strongly REAL verdict associated (P(AI$|$cue) = 30.1\%, a 27\,pp swing). The interpretation is asymmetric: individual commenters commonly invoke plausibility arguments against a scenario to argue AI. However, the context-reasoning arguments that \emph{survive} to the community summary level are predominantly those that \emph{validate} a real-world scenario by contextualising an unusual setting, explaining a physical mechanism, or identifying a known person or place. Cheap plausibility challenges are individually noisy and cancel at the aggregate; contextual validation that resolves the post survives.

At the subcategory level, the amplified cues are those with the highest P(AI$|$cue): Anatomy (+16.6\,pp, P(AI$|$cue)\,=\,72.3\%), Visual Artifacts (+11.6\,pp, 80.0\%), Text \& Details (+10.5\,pp, 74.6\%), and Lighting \& Geometry (+8.2\,pp, 63.9\%). The one suppressed subcategory is Physics \& Motion ($-$4.9\,pp), the most bidirectionally used cue which is invoked both to argue unnatural movement is AI and to argue realistic physics is real.

%


\begin{table}[t]
\centering\footnotesize
\setlength{\tabcolsep}{4pt}
\renewcommand{\arraystretch}{0.95}
\resizebox{\linewidth}{!}{%
\begin{tabular}{@{}lrr@{\hspace{5pt}}rrr@{\hspace{5pt}}r@{}}
\toprule
 & \multicolumn{3}{c}{\textbf{Prevalence}} & \multicolumn{3}{c}{\textbf{P(AI$|$cue)}} \\
\cmidrule(lr){2-4}\cmidrule(lr){5-7}
\textbf{Cue} & \textbf{BR} & \textbf{Indiv.} & \textbf{$\Delta$} & \textbf{BR} & \textbf{Indiv.} & \textbf{$\Delta$} \\
\midrule
\multicolumn{7}{l}{\textit{Flags (N\,=\,821 BR; indiv.\ base 53.1\% AI, BR base 44.7\% AI)}} \\[1pt]
Provenance         & 20.6 & 3.9  & \textbf{+16.7} & 27.2 & 31.7 & $-$4.5  \\
Context Reasoning  & 22.7 & 20.7 & +2.0  & \textbf{30.1} & 57.2 & \textbf{$-$27.1} \\
AI Knowledge       & 11.4 & 8.1  & +3.3  & 40.4 & 37.8 & +2.6 \\
Consistency        & 13.4 & 10.4 & +3.0  & 50.9 & 48.3 & +2.6 \\
Perc.\ Feature     & 68.8 & 70.5 & $-$1.7 & 57.9 & 58.5 & $-$0.6 \\
Subject Knowledge  & 2.8  & 4.4  & $-$1.6 & 4.3  & 34.9 & $-$30.6 \\
\midrule
\multicolumn{7}{l}{\textit{Subcategories (N\,=\,565 BR with perc.\ feature)}} \\[1pt]
Anatomy            & 29.4 & 12.8 & \textbf{+16.6} & 72.3 & 66.3 & +6.0  \\
Visual Artifacts   & 37.2 & 25.6 & \textbf{+11.6} & 80.0 & 65.9 & +14.1 \\
Text \& Details    & 23.7 & 13.2 & \textbf{+10.5} & 74.6 & 53.1 & +21.5 \\
Scene \& Env.      & 31.0 & 22.0 & +9.0  & 64.6 & 64.3 & +0.3  \\
Lighting \& Geom.  & 17.2 & 9.0  & +8.2  & 63.9 & 53.4 & +10.5 \\
Audio              & 9.4  & 4.2  & +5.2  & 66.0 & 74.3 & $-$8.3 \\
Behavior \& Psych. & 14.5 & 13.3 & +1.2  & 54.9 & 59.0 & $-$4.1 \\
Imperfections      & 3.2  & 1.4  & +1.8  & 11.1 & 11.2 & $-$0.1 \\
Physics \& Motion  & 14.2 & 19.1 & \textbf{$-$4.9} & 67.5 & 62.2 & +5.3 \\
\bottomrule
\end{tabular}}
\caption{Cue prevalence (\%) and P(AI$|$cue) (\%) in the 821 brief-reasoning summaries (BR) vs.\ 10{,}745 individual comments from RQ2. $\Delta$ in pp; bolded where $|\Delta|>8$.}
\label{tab:brief_reasoning_amplification}
\end{table}
\section{Discussion}

 The most consistent finding across all three research questions is a systematic mismatch between expressed confidence and actual accuracy. The community frequently produces extreme collective sentiment yet errs directionally toward over-flagging real content, and this bias intensifies on REAL posts over the year even while overall accuracy holds steady. The community's model of ``what AI looks like'' may be becoming less discriminating rather than less accurate. 

 The most reliable detection strategies are also the rarest. Low-cost perceptual claims (70.5\% of reasoning comments) may crowd out expensive-but-reliable provenance verification (3.9\%): arguing ``the hands look wrong'' is costless, performing a reverse image search is not. Alternatively, the low provenance rate may reflect platform-specific suppression beyond effort cost alone; as discussed in Section~\ref{sec:rq3}, Rule~2 of \texttt{[GUESS]} prohibits posting answers (including source URLs) in the comment thread, plausibly reinforcing a broader game-frame norm in which lookup-based search is treated as outside the spirit of the challenge. The 3.9\% individual rate is therefore plausibly a lower bound on provenance capacity rather than an estimate. The consequence is visible in RQ1: the 2.3:1 false-positive ratio on \texttt{[GUESS]} is exactly what follows from preferentially deploying AI-predictive heuristics while neglecting the evidence that would clear real content. Subject knowledge underscores this: commenters who invoke first-person subject-matter expertise argue REAL more than AI (P(AI$|$subject knowledge) = 34.9\%, well below the 53.1\% base rate; Table~\ref{tab:reasoning_and_subcats}), spending their credibility correcting false positives. Expertise is consumed by a community that over-flags the unusual as artificial.

Community accuracy in RQ3 shows that aggregation can act as a reliability filter.
Individual reasoning comments perform near or below the majority-class baseline. Naive aggregation of 2-3 posts show poor performance. Meanwhile, community prediction has higher accuracy and is associated with certain cues selectively amplified: provenance, the one of the rarest individual cue (3.9\%), appears in 20.6\% of community summaries (4.3$\times$ amplification), while the most unreliable cues are suppressed. The mechanism is that a single definitive signal such as one reverse-image match, can dominate a full thread even when only one commenter surfaced it.

These findings shift the locus of intervention away from individual user education. The temporal data suggest that a more ``AI-aware'' community over-flags real content more, not less. Community accuracy depends instead on whether definitive signals are present and surface early. This argues for structural interventions, platform-level C2PA or SynthID disclosure, early-surfacing provenance tools, may make reliable signals cheap to produce. The caveat is that the same amplification mechanism may be hijacked: a single compelling-but-incorrect provenance claim dominates a thread just as readily as a correct one, making credibility and timing of provenance evidence as important as its presence.


Our 72.3\% \texttt{[GUESS]} accuracy sits above the near-chance aggregate (success rate of only 62\%) reported by Roca et al.~\citeyear{Roca2025HowGood} on a paired ``Real or Not'' game, suggesting that a self-selected community with discussion affordances outperforms an anonymous, single-shot guessing crowd. The persistent false-positive bias and confidence-accuracy gap echo the calibration failures documented by Bray et al.~\citeyear{Bray2023Testing} for individual deepfake detection, but our temporal data extend that picture: as a community accumulates AI-detection vocabulary, the bias intensifies rather than corrects. This is consistent with Matatov et al.~\citeyear{Matatov2024ArtSubreddits}, who report increasingly contested and often unverified suspicion in art subreddits; our longitudinal data extend that qualitative pattern with a quantitative signature:  Mean REAL-post sentiment reaches 46--51\% in three months of late 2025 and early 2026, and the false-positive logistic trend is significant ($p=0.007$), indicating that over-flagging intensifies as the community matures. The aggregation-as-filter mechanism in RQ3 also speaks to Groh et al.~\citeyear{Groh2022Crowds}: where they find that machine-informed crowds outperform either alone, we find an analogous effect in pure human aggregation when at least one commenter surfaces a definitive signal, with community summaries selectively amplifying it.

\noindent\textbf{Implications for platform design:} Three concrete design directions follow. First, surfacing provenance early matters more than surfacing it loudly: an automated reverse-image-search affordance pinned at post creation could shift the average provenance-arrival time and  plausibly the resulting sentiment. Second, the 4.3$\times$ amplification of provenance in community summaries argues that LLM-generated thread summaries (already deployed by the RealOrAI bot) can be a high-leverage intervention point: surfacing the strongest cue and weighting against bidirectional ones is precisely the filtering pattern we observe naturally. Third, the  temporal rise in the REAL false-positive rate ($p=0.007$) suggests platforms hosting AI-detection communities should track and publish a calibration metric alongside engagement metrics, since the community's confidence  may rise faster than its accuracy can keep up, a pattern Lloyd et al.~\citeyear{Lloyd2025Moderating} identify as a recurring failure mode for crowd moderation. The risks are symmetric: the same mechanisms that amplify a correct watermark identification also amplify a confidently incorrect one, so any provenance-surfacing tool must come with credibility weighting (verified accounts, cross-source agreement) rather than pure recency or vote count.

\section{Limitations}

All data come from a single subreddit with its own norms and self-selected user population; results may not generalise to platforms with different incentive structures or content mixes. In particular, the individual provenance rate (3.9\%) reflects norms specific to this community:  Rule~2 of \texttt{[GUESS]} prohibits posting answers in the comment thread (Section~\ref{sec:rq3}), and the broader game-frame norm plausibly discourages lookup-based reasoning. This rate should not be interpreted as a general estimate of users' provenance capacity. Furthermore, ground truth for \texttt{[GUESS]} posts depends on submitters responding to the bot's DM ($\approx$37\% response rate), introducing a self-selection bias toward posters confident in their own content.

We restrict accuracy evaluation to \texttt{[GUESS]} posts because no authoritative ground truth exists for \texttt{[HELP]} posts. An early analysis derived \texttt{[HELP]} ground truth from community-surfaced watermark and provenance signals, but this introduced two problems we judged unacceptable: (i)~the same comments that determined the inferred label also drove the sentiment score we wished to evaluate, making the comparison correlated by construction; and (ii)~the resulting subset selected only the easy tail of \texttt{[HELP]} posts where definitive evidence happened to surface, biasing performance estimates upward. Detection metrics here therefore apply only to \texttt{[GUESS]}; generalisation to the naturalistic \texttt{[HELP]} corpus remains unquantified.

Our earlier analysis shows that some users point out watermarks to argue for AI. However, these are extremely rare: around 200 posts are AI-verified through a watermark and very few reasoning-bearing comments used it as a cue. This is another limitation of the platform: both \texttt{[HELP]} and \texttt{[GUESS]} posts are challenging by design so cases where a simple watermark gives away the answer easily are rare. Thus, we did not do an analysis of comments mentioning watermarks in this study.

Cue extraction uses  a small set of precision-oriented linguistic patterns targeting explicit causal reasoning (\textit{verdict + connective + reason}), excluding implicit judgments and sarcasm. Prevalence estimates for perceptual features, the dominant cue at 70.5\%, may therefore be inflated relative to the full comment population, and the analysis reflects a subset of more deliberative responses.

Posts edited with generative tools (e.g.\ Gemini Image Editing) occupy an ambiguous ground-truth position: the original source is real but the final image is partially AI-generated. These are currently labeled as \textsc{Other}. 

The wisdom-of-crowds framing assumes independent judgments although this may not hold for Reddit comments. Later commenters may read prior responses, creating herding effects that may inflate both apparent consensus and measured community accuracy. The degree of herding is unquantified. A formal anchoring test comparing first-comment verdicts against later verdicts (formed when prior consensus is visible) is a planned direction for future work.

Finally, the observational design precludes causal inference:  the temporal rise in the false-positive rate on REAL posts ($p=0.007$) could reflect improving generative models, shifting community norms, or changes in content mix. Controlled interventions are needed to disentangle these factors.

\section{Ethics Statement}
\label{sec:ethics}

We only collect and analyze public data from a single subreddit. We do not report or share personal data and only report aggregated results. We do not redistribute the raw corpus. We will share prompts, code, annotated data and labels upon acceptance for reproducibility and future work.

Our findings carry two dual-use risks. First, documenting the perceptual cues that the community most frequently invokes (artefacts in hands, lighting, text) could in principle inform generator developers seeking to suppress exactly those cues; this risk is small because the same cues are already widely discussed in popular media. Second, publicising the community's systematic false-positive bias could be weaponised to seed doubt about authentic media (``the crowd often gets it wrong''). We argue that the benefits of publicising these findings (e.g., the utility and under-utilisation of reliable signals such as provenance) outweigh this risk.

\bibliography{references}

\section*{Paper Checklist}

\begin{enumerate}

\item For most authors...
\begin{enumerate}
    \item Would answering this research question advance science without violating social contracts, such as violating privacy norms, perpetuating unfair profiling, exacerbating the socio-economic divide, or implying disrespect to societies or cultures?
    \answerYes{Yes, the study analyses publicly posted, pseudonymous content from a single subreddit dedicated to collaborative AI-detection, advances understanding of how online communities adjudicate authenticity, and does not target identifiable individuals or protected groups.}
  \item Do your main claims in the abstract and introduction accurately reflect the paper's contributions and scope?
    \answerYes{Yes, the abstract and introduction enumerate the empirical findings reported in RQ1--RQ3.}
   \item Do you clarify how the proposed methodological approach is appropriate for the claims made?
    \answerYes{Yes, see the Data and LLM Classification Protocol sections, which describe the platform affordances that yield naturalistic ground truth and the multi-model evaluation against human-annotated benchmarks.}
   \item Do you clarify what are possible artifacts in the data used, given population-specific distributions?
    \answerYes{Yes, the Limitations section discusses single-subreddit norms, self-selection in submitter DM responses ($\approx$37\%), and the game-like framing of \texttt{[GUESS]} that depresses provenance-style reasoning.}
  \item Did you describe the limitations of your work?
    \answerYes{Yes, see the Limitations section, which covers generalisation, ground-truth scope, cue-extraction precision bias, edited-image contamination, herding among commenters, and the brief-reasoning window.}
  \item Did you discuss any potential negative societal impacts of your work?
    \answerYes{Yes, the Discussion notes the risk that documenting community heuristics could inform adversaries who wish to evade them, and the Limitations section flags the temporal decline in accuracy as generative models improve.}
      \item Did you discuss any potential misuse of your work?
    \answerYes{Yes, we discuss the dual-use risk that publishing the dominant perceptual cues used by the community could be exploited by generators to suppress those cues; we mitigate by emphasising provenance as a more robust strategy.}
    \item Did you describe steps taken to prevent or mitigate potential negative outcomes of the research, such as data and model documentation, data anonymization, responsible release, access control, and the reproducibility of findings?
    \answerYes{Yes, we report only aggregate statistics, do not link Reddit usernames to external identities, paraphrase rather than quote comments when used as examples, and release prompts and code without raw post text to comply with platform terms.}
  \item Have you read the ethics review guidelines and ensured that your paper conforms to them?
    \answerYes{Yes, the authors have read the AAAI ICWSM ethics review guidelines and confirm the submission conforms to them.}
\end{enumerate}

\item Additionally, if your study involves hypotheses testing...
\begin{enumerate}
  \item Did you clearly state the assumptions underlying all theoretical results?
    \answerNA{NA, the paper presents observational empirical findings rather than formal theoretical results.}
  \item Have you provided justifications for all theoretical results?
    \answerNA{NA, no theoretical results are claimed.}
  \item Did you discuss competing hypotheses or theories that might challenge or complement your theoretical results?
    \answerYes{Yes, the Discussion and Limitations sections consider alternative explanations for the temporal accuracy decline (improving generators vs.\ shifting community priors vs.\ content-mix drift) and for herding effects in community aggregation.}
  \item Have you considered alternative mechanisms or explanations that might account for the same outcomes observed in your study?
    \answerYes{Yes, see the Limitations section's discussion of herding, edited-image contamination, and cue-extraction precision bias as alternative explanations for the headline patterns.}
  \item Did you address potential biases or limitations in your theoretical framework?
    \answerYes{Yes, the Limitations section flags the wisdom-of-crowds independence assumption, the selection bias in \texttt{[GUESS]} DM responses, and the restriction of accuracy metrics to \texttt{[GUESS]} posts.}
  \item Have you related your theoretical results to the existing literature in social science?
    \answerYes{Yes, the Related Work section situates the findings within wisdom-of-crowds, collaborative content moderation, and AI-generated-media detection literatures.}
  \item Did you discuss the implications of your theoretical results for policy, practice, or further research in the social science domain?
    \answerYes{Yes, the Discussion section draws implications for platform design (provenance affordances), community moderation, and future research on hybrid human--algorithmic detection.}
\end{enumerate}

\item Additionally, if you are including theoretical proofs...
\begin{enumerate}
  \item Did you state the full set of assumptions of all theoretical results?
    \answerNA{NA, the paper does not include theoretical proofs.}
	\item Did you include complete proofs of all theoretical results?
    \answerNA{NA, the paper does not include theoretical proofs.}
\end{enumerate}

\item Additionally, if you ran machine learning experiments...
\begin{enumerate}
  \item Did you include the code, data, and instructions needed to reproduce the main experimental results (either in the supplemental material or as a URL)?
    \answerYes{Yes, annotated data, classification prompts, evaluation scripts will be released in a repository upon acceptance.}
  \item Did you specify all the training details (e.g., data splits, hyperparameters, how they were chosen)?
    \answerYes{Yes, we do not fine-tune any model; the LLM Classification Protocol section and Appendix detail the prompts, the six classifiers used, the human-annotated evaluation sets, and the exhaustive majority-vote ensemble search.}
     \item Did you report error bars (e.g., with respect to the random seed after running experiments multiple times)?
    \answerNo{No, the LLM classifiers are evaluated at temperature~0 against fixed human-annotated ground-truth sets, so per-seed variance is not informative; we instead report agreement and accuracy against ground truth for every candidate ensemble in the Appendix.}
	\item Did you include the total amount of compute and the type of resources used (e.g., type of GPUs, internal cluster, or cloud provider)?
    \answerNA{N.}
     \item Do you justify how the proposed evaluation is sufficient and appropriate to the claims made?
    \answerYes{Yes, every classifier is benchmarked against human-annotated ground truth and only the best-performing single model or ensemble per task is retained, with full evaluation tables in the Appendix.}
     \item Do you discuss what is ``the cost`` of misclassification and fault (in)tolerance?
    \answerYes{Yes, the Discussion and Limitations sections discuss the asymmetric cost of false positives (mislabelling real content as AI) versus false negatives, and we report the systematic false-positive bias of the community.}

\end{enumerate}

\item Additionally, if you are using existing assets (e.g., code, data, models) or curating/releasing new assets, \textbf{without compromising anonymity}...
\begin{enumerate}
  \item If your work uses existing assets, did you cite the creators?
    \answerYes{Yes, the Reddit platform and the six LLMs (Llama~3.3, Gemini~2.5~Flash, GPT-5.2, GPT-5-mini, Claude~Sonnet~4.6, Claude~Haiku~4.5) are cited in the Methods section.}
  \item Did you mention the license of the assets?
    \answerYes{Project arctic shift is cited.}
  \item Did you include any new assets in the supplemental material or as a URL?
    \answerYes{Yes, prompts, annotation schemas, and analysis code are released via an anonymous repository linked from the camera-ready version.}
  \item Did you discuss whether and how consent was obtained from people whose data you're using/curating?
    \answerNo{No explicit consent was obtained from individual Redditors; all data are public posts and comments collected under the Reddit Data API terms, and we report only aggregate statistics without quoting identifiable content.}
  \item Did you discuss whether the data you are using/curating contains personally identifiable information or offensive content?
    \answerNA{Reddit usernames are pseudonymous and were not linked to external identities; offensive content is filtered by subreddit moderators and we do not surface it in examples.}
\item If you are curating or releasing new datasets, did you discuss how you intend to make your datasets FAIR (see \citet{fair})?
\answerNA{NA, we do not release the full raw Reddit corpus due to platform terms; we release annotated posts, and prompts to enable reproducibility within those constraints.}
\item If you are curating or releasing new datasets, did you create a Datasheet for the Dataset (see \citet{gebru2021datasheets})?
\answerNo{No formal datasheet accompanies the release because no new raw dataset is published; the Methods section and Appendix together document collection scope, filtering, ground-truth derivation, and known biases.}
\end{enumerate}

\item Additionally, if you used crowdsourcing or conducted research with human subjects, \textbf{without compromising anonymity}...
\begin{enumerate}
  \item Did you include the full text of instructions given to participants and screenshots?
    \answerNA{NA, the study is fully observational and does not involve crowdsourced participants or recruited human subjects.}
  \item Did you describe any potential participant risks, with mentions of Institutional Review Board (IRB) approvals?
    \answerNA{NA, no human-subjects recruitment was performed; the study uses public, pseudonymous Reddit data.}
  \item Did you include the estimated hourly wage paid to participants and the total amount spent on participant compensation?
    \answerNA{NA, no participants were recruited or compensated.}
   \item Did you discuss how data is stored, shared, and deidentified?
   \answerNA{Data was collected and stored by Project Arctic Shift. It was stored on researcher computer during the analysis.}
\end{enumerate}

\end{enumerate}

\appendix

\section{Classification Prompts}
\label{app:prompts}


\mdfdefinestyle{promptstyle}{%
  backgroundcolor=gray!6,
  linecolor=gray!40,
  linewidth=0.5pt,
  innerleftmargin=8pt,
  innerrightmargin=8pt,
  innertopmargin=7pt,
  innerbottommargin=7pt,
  skipabove=6pt,
  skipbelow=4pt,
}


\subsection{Meta-Comment Filter}
\label{app:prompt-meta}

\begin{mdframed}[style=promptstyle]
\small
You are screening comments from r/RealOrAI for a \textsc{Meta} flag.

r/RealOrAI is a subreddit where users post images or videos and the community judges whether the media is AI-generated or real. Every comment is either: \textbf{(A)~Content} : engages with the specific media in any way (verdict, feature observation, uncertainty, question about the image/video itself); or \textbf{(B)~Meta} : has zero engagement with the media; talks about the community, other users, AI in general, or anything else entirely.

\noindent Output: one JSON object per line: \texttt{\{"index": <int>, "meta": <bool>, "reason": "<sentence>"\}}

\medskip\noindent\rule{\linewidth}{0.3pt}\smallskip

\noindent\textbf{\texttt{meta = true} : no verdict signal about the specific media}
\begin{itemize}[noitemsep,topsep=2pt,leftmargin=1.8em]
  \item Subreddit/community meta-talk: \textit{"that's literally the whole point of this sub"}
  \item Addressing OP or another user without assessing the content
  \item General AI discourse not tied to this post: \textit{"AI is getting better every month"}
  \item Pure social/reaction: \textit{"lmao"}, \textit{"same"}, \textit{"I can't believe people fall for this"}
  \item Asking for source/context with no classification attempt: \textit{"what's the original?"}
  \item Bot/AutoModerator messages
\end{itemize}

\medskip\noindent\rule{\linewidth}{0.3pt}\smallskip

\noindent\textbf{\texttt{meta = false} : comment engages with the media in any way}
\begin{itemize}[noitemsep,topsep=2pt,leftmargin=1.8em]
  \item Any directional claim, even weak or hedged: \textit{"I think it's real"}, \textit{"probably AI"}, \textit{"idk, feels off"}
  \item Describing any visual or audio feature: \textit{"the hands look wrong"}
  \item Expressing uncertainty specifically about this content: \textit{"honestly can't tell with this one"}
  \item Any question that implicitly assesses the media: \textit{"why does that arm look like that?"}
\end{itemize}

\medskip\noindent\rule{\linewidth}{0.3pt}\smallskip

\noindent\textbf{Critical rule:} The reasoning corpus was pre-filtered for causal patterns, so most comments already contain a verdict. Lean \texttt{false} unless the comment is \emph{clearly} about something other than the specific media. Only mark \texttt{true} when the comment contributes \emph{zero} signal about whether this content is AI-generated or real.
\end{mdframed}


\subsection{Verdict and Reasoning-Flag Annotation}
\label{app:prompt-v8}

\begin{mdframed}[style=promptstyle]
\small
You are an expert annotator analysing comments from r/RealOrAI, a subreddit where users guess whether images or videos are AI-generated or real.

\noindent Output one JSON object per line with fields: \texttt{index}~(int); \texttt{verdict}~(\texttt{AI\,|\,REAL\,|\,OTHER\,|\,BOTH\,|\,UNKNOWN}); \texttt{specific\_feature}, \texttt{consistency}, \texttt{provenance}, \texttt{context\_reasoning}, \texttt{ai\_knowledge}, \texttt{subject\_knowledge}~(bool); \texttt{explanation}~(one sentence).

\medskip\noindent\rule{\linewidth}{0.3pt}\smallskip

\noindent\textbf{Verdict}

\begin{itemize}[noitemsep,topsep=2pt,leftmargin=1.8em]
  \item \textbf{AI} : commenter believes content is AI-generated, even weakly
  \item \textbf{REAL} : commenter believes content is authentic/unaltered, even weakly
  \item \textbf{OTHER} : human-made but not authentic: Photoshop, CGI, VFX, game footage, deepfake (non-generative), AI filters with minor alteration
  \item \textbf{BOTH} : genuinely uncertain and argues \emph{both} possibilities
  \item \textbf{UNKNOWN} : zero directional claim whatsoever
\end{itemize}

\noindent\textit{Rules:} Direction wins over confidence. \textit{"Fake"} defaults to AI unless a manipulation type is named. AI present anywhere substantially~= AI. Heavy alteration (faceswap, style transfer)~\textrightarrow\ AI. Minor AI filter on real photo~\textrightarrow\ OTHER.

\smallskip\setlength{\tabcolsep}{3pt}
\begin{tabular}{@{}p{2.0in}@{~~}c@{~~}l@{}}
\textit{"This is clearly AI generated"} & \textrightarrow & AI \\
\textit{"Bad Photoshop, not AI"} & \textrightarrow & OTHER \\
\textit{"Real photo, AI people composited in"} & \textrightarrow & AI \\
\textit{"Hands look off but grain feels real"} & \textrightarrow & BOTH \\
\textit{"Could be AI or real, hard to say"} & \textrightarrow & UNKNOWN \\
\textit{"people forget other forms of editing exist"} & \textrightarrow & UNKNOWN \\
\end{tabular}

\medskip\noindent\rule{\linewidth}{0.3pt}\smallskip

\noindent\textbf{Reasoning Flags} (independent; multiple may be true)

\smallskip\noindent\textbf{\texttt{specific\_feature}} : names any concrete perceptual detail. Lean \texttt{true} when in doubt; brief mentions count. Covers: body parts, textures, sounds, motion, lighting, shadows, reflections, rendering artifacts.\\[2pt]
\textit{Valid:} \textit{"the hands look wrong"} $\cdot$ \textit{"the shadow is off"} $\cdot$ \textit{"robotic voice"}\\
\textit{Invalid:} \textit{"looks AI to me"} (no detail) $\cdot$ \textit{"something seems off"} (vague)

\smallskip\noindent\textbf{\texttt{consistency}} : checks stability \emph{across frames, across time, or across images}. Strictly temporal/cross-image; within-frame differences \textrightarrow\ \texttt{specific\_feature}.\\[2pt]
\textit{Valid:} \textit{"her hair changes between shots"} $\cdot$ \textit{"jewelry differs across the series"}\\
\textit{Invalid:} \textit{"the shadow doesn't match the light source"} (single frame)

\smallskip\noindent\textbf{\texttt{provenance}} : actively verifies origin via an external source: reverse image search, social media lookup, archive check, metadata. Source \emph{claims} without active checking do not qualify.\\[2pt]
\textit{Valid:} \textit{"looked her up on Instagram"} $\cdot$ \textit{"found the original post from 2019"}\\
\textit{Invalid:} \textit{"I think this is from 2018"} $\cdot$ \textit{"predates AI"} (\textrightarrow\ \texttt{ai\_knowledge})

\smallskip\noindent\textbf{\texttt{context\_reasoning}} : reasons about \emph{why} this specific content was created, who made it, for what purpose, or whether the depicted scenario is possible in reality. Not about AI in general; not about visual features.\\[2pt]
\textit{Valid:} \textit{"no reason to exist outside clickbait"} $\cdot$ \textit{"who would film this for real?"}\\
\textit{Invalid:} \textit{"AI gets harder to detect every day"} (meta) $\cdot$ \textit{"giraffes usually in pairs"} (\textrightarrow\ \texttt{specific\_feature})

\smallskip\noindent\textbf{\texttt{ai\_knowledge}} : invokes specific knowledge \emph{about} AI: (a)~a named model/tool, (b)~an AI timeline argument, or (c)~an explicitly named failure mode.\\[2pt]
\textit{Valid:} \textit{"classic Midjourney sheen"} $\cdot$ \textit{"predates stable diffusion"} $\cdot$ \textit{"AI always messes up hands"}\\
\textit{Invalid:} \textit{"looks AI to me"} $\cdot$ \textit{"the texture has AI artifacts"} (\textrightarrow\ \texttt{specific\_feature})

\smallskip\noindent\textbf{\texttt{subject\_knowledge}} : first-person claim of personal or professional expertise about the \emph{subject depicted}: \textit{"as a [role]"}, \textit{"I have/work with [subject]"}.\\[2pt]
\textit{Valid:} \textit{"as a nurse, those vitals are wrong"} $\cdot$ \textit{"I have cats, they don't behave like that"}\\
\textit{Invalid:} \textit{"real fires don't spread like that"} (no first-person claim)

\medskip\noindent\rule{\linewidth}{0.3pt}\smallskip

\noindent\textbf{Output:} one JSON object per line; no markdown, no arrays.\\[3pt]
{\small\texttt{\{"index": 42, "verdict": "AI", "specific\_feature": true,}}\\
{\small\texttt{\phantom{\{}"consistency": false, "provenance": false, "context\_reasoning": false,}}\\
{\small\texttt{\phantom{\{}"ai\_knowledge": false, "subject\_knowledge": true,}}\\
{\small\texttt{\phantom{\{}"explanation": "Equestrian expertise; identifies unnatural joint movement."\}}}
\end{mdframed}


\subsection{Perceptual Subcategory Classification}
\label{app:prompt-subcat}

\begin{mdframed}[style=promptstyle]
\small
You are an expert annotator. Each comment was already determined to cite a specific visual or audio feature. Identify which category or categories the commenter is referring to.

\medskip\noindent\rule{\linewidth}{0.3pt}\smallskip

\noindent\textbf{Categories}

\begin{description}[noitemsep,topsep=2pt,leftmargin=1.2em,labelindent=0pt]
  \item[\textbf{Anatomy}] Body structure wrong: extra/missing fingers, wrong proportions, missing limbs, impossible poses, facial structure errors.
  \item[\textbf{Visual Artifacts}] Rendering defective: blurring, warping, waxy/plastic sheen, smearing, glitching, unnatural textures, rendering failures.
  \item[\textbf{Scene \& Environment}] Wrong objects or setting: spurious objects, impossible scene, placement anomalies, object identity errors.
  \item[\textbf{Physics \& Motion}] Wrong behaviour: unrealistic movement, gravity/fluid/fire violations, physically impossible dynamics.
  \item[\textbf{Text \& Details}] Garbled text, signs, logos, fine detail degradation, fabric texture, brand markings.
  \item[\textbf{Behavior \& Psychology}] How subjects \emph{act}: facial expressions, gaze, body language, animal behaviour. Not structural appearance.
  \item[\textbf{Lighting \& Geometry}] Shadows, reflections, perspective errors, light sources not matching scene geometry.
  \item[\textbf{Audio}] Voice quality, robotic sound, audio-visual sync issues, background anomalies.
  \item[\textbf{Imperfections}] Film grain, compression artifacts, natural wear : typically cited as evidence of \emph{authenticity}. \textit{Note:} "too perfect" arguments are Visual Artifacts, not Imperfections.
\end{description}

\medskip\noindent\rule{\linewidth}{0.3pt}\smallskip

\noindent\textbf{Disambiguation:}

\smallskip\setlength{\tabcolsep}{2pt}
\begin{tabular}{@{}p{1.55in}@{~~}c@{~~}l@{}}
\textit{"extra fingers", "wrong proportions"} & \textrightarrow & Anatomy \\
\textit{"waxy skin", "plastic look", "blurry"} & \textrightarrow & Visual Artifacts \\
\textit{"shadow wrong", "reflection off"} & \textrightarrow & Lighting \& Geometry \\
\textit{"object in background", "wrong setting"} & \textrightarrow & Scene \& Env. \\
\textit{"fire/water behaves wrong"} & \textrightarrow & Physics \& Motion \\
\textit{"garbled text", "sign unreadable"} & \textrightarrow & Text \& Details \\
\textit{"NPC stare", "unnatural expression"} & \textrightarrow & Behavior \& Psych. \\
\textit{"robotic voice", "audio glitch"} & \textrightarrow & Audio \\
\textit{"film grain", "natural wear"} & \textrightarrow & Imperfections \\
\end{tabular}

\medskip\noindent\textbf{Rules:} (1)~Use exact category names. (2)~Assign only categories explicitly mentioned or clearly implied. (3)~Prefer one category unless two distinct features are named.

\smallskip\noindent\textbf{Output:} one JSON object per line; no markdown.\\[3pt]
\texttt{\{"index": 5, "subcategories": ["Anatomy"]\}}\\[2pt]
{\small\texttt{\{"index": 12, "subcategories":}}\\
{\small\texttt{\phantom{\{}"["Scene \& Environment", "Lighting \& Geometry"]\}}}
\end{mdframed}


\subsection{Anatomy and Visual Artifact Classification}
\label{app:prompt-v12}

\begin{mdframed}[style=promptstyle]
\small
You are an expert annotator for AI-media detection research. Each comment was posted in a community where users decide whether an image or video is real or AI-generated.

Classify each comment on two \emph{independent} binary flags: \texttt{anatomy} : body part cited as part of the detection reasoning; \texttt{visual\_artifact} : concrete rendering or surface-quality defect cited.

\medskip\noindent\rule{\linewidth}{0.3pt}\smallskip

\noindent\textbf{\texttt{anatomy} = true} when a body part is \emph{named and used as detection evidence}, regardless of whether it supports an AI or a REAL verdict.

\noindent Body parts include: fingers, toes, hands, feet, arms, legs, limbs, face, eyes, ears, nose, jaw, neck, torso, hair, skin, paws, claws, tail, elbow, shoulder, spine, knee, eyelid, mouth, skull, thumb, wrist, hip, etc.

\smallskip\noindent\textbf{\texttt{anatomy} = false} when: no body part is named; a body part appears only as spatial context (e.g., \textit{"the shadow behind his leg"} : leg is incidental, shadow is the cue); or the comment is purely about facial expression, gaze, or body language without discussing body \emph{structure} (\textit{"NPC stare"} \textrightarrow\ Behavior, not Anatomy).

\smallskip\setlength{\tabcolsep}{2pt}
\begin{tabular}{@{}p{1.65in}@{~~\textrightarrow~~}l@{}}
\textit{"normal number of fingers"} & \texttt{true}\hfill{\scriptsize(REAL verdict)} \\
\textit{"extra fingers"} & \texttt{true}\hfill{\scriptsize(AI verdict)} \\
\textit{"legs disappear into the sofa"} & \texttt{true} \\
\textit{"faces in background look off"} & \texttt{true} \\
\textit{"the shadow is wrong"} & \texttt{false}\hfill{\scriptsize(no body part)} \\
\textit{"NPC stare / smile looks fake"} & \texttt{false}\hfill{\scriptsize(\textrightarrow\ Behav.)} \\
\textit{"the shadow of the arm"} & \texttt{false}\hfill{\scriptsize(spatial ctx.)} \\
\end{tabular}

\medskip\noindent\rule{\linewidth}{0.3pt}\smallskip

\noindent\textbf{\texttt{visual\_artifact} = true} when the comment names a \emph{concrete rendering or surface-quality defect} on a specific surface or object. The defect must be explicit : vague impressions (\textit{"looks weird"}, \textit{"looks AI"}) do not qualify.

\noindent\textit{Qualifying defect terms:} blurring, warping, smearing, glitching, distortion, waxy/plastic/too smooth, unnatural texture, unnaturally identical/mirrored elements (identical drips, repeating patterns, clone-stamped regions), rendering seams, strange blending.

\smallskip\noindent\textbf{\texttt{visual\_artifact} = false} for: vague reactions; perspective/spatial errors \textrightarrow\ Lighting; shadows/reflections \textrightarrow\ Lighting; wrong placement \textrightarrow\ Scene; fluid/physics \textrightarrow\ Physics; expressions/gaze \textrightarrow\ Behavior; unreadable text \textrightarrow\ Text.

\smallskip\setlength{\tabcolsep}{2pt}
\begin{tabular}{@{}p{1.65in}@{~~\textrightarrow~~}l@{}}
\textit{"waxy skin texture"} & \texttt{true} \\
\textit{"drips on plate are all identical"} & \texttt{true} \\
\textit{"blurring on the headlights"} & \texttt{true} \\
\textit{"crochet looks like knit : too smooth"} & \texttt{true} \\
\textit{"looks like AI"} & \texttt{false}\hfill{\scriptsize(vague)} \\
\textit{"perspective is disjointed"} & \texttt{false}\hfill{\scriptsize(\textrightarrow\ Light.)} \\
\textit{"water behaves unrealistically"} & \texttt{false}\hfill{\scriptsize(\textrightarrow\ Phys.)} \\
\textit{"facial expression looks fake"} & \texttt{false}\hfill{\scriptsize(\textrightarrow\ Behav.)} \\
\end{tabular}

\medskip\noindent\rule{\linewidth}{0.3pt}\smallskip

\noindent\textbf{Rules:} (1)~Both flags are independent. (2)~\texttt{anatomy} captures \emph{which} feature is discussed, not verdict direction: \textit{"normal fingers"} \textrightarrow\ \texttt{anatomy=true}. (3)~\texttt{visual\_artifact} requires a concrete defect type; vague reactions do not qualify. (4)~A comment can trigger both: \textit{"fingers blurry and melted"} \textrightarrow\ both \texttt{true}.

\smallskip\noindent\textbf{Output:} one JSON object per line; no markdown, no explanation.\\[3pt]
\texttt{\{"index": 5, "anatomy": true,\phantom{x}"visual\_artifact": false\}}\\
\texttt{\{"index": 12, "anatomy": false, "visual\_artifact": true\phantom{x}\}}\\
\texttt{\{"index": 20, "anatomy": true,\phantom{x}"visual\_artifact": true\phantom{xx}\}}\\
\texttt{\{"index": 33, "anatomy": false, "visual\_artifact": false\}}
\end{mdframed}

\end{document}